\documentclass{article}
\usepackage{spconf,amsmath,graphicx}

\usepackage{enumitem}

\setlist{nosep, leftmargin=14pt}

\usepackage{mwe} 
\usepackage{color, colortbl}
\definecolor{Gray}{gray}{0.9}
\usepackage{graphicx}
\usepackage{xcolor}
\usepackage{makecell}
\usepackage{url}

\usepackage{amsmath}
\usepackage{amssymb}
\usepackage{lipsum}
\usepackage{subfig}
\usepackage{booktabs}
\usepackage{pifont}
\newcommand{\cmark}{\ding{51}}%
\newcommand{\xmark}{\ding{55}}%

\usepackage{float}
\floatstyle{plaintop}
\restylefloat{table}

\usepackage[acronym,nomain]{glossaries}

\newacronym{mr}{MR}{Magnetic Resonance}
\newacronym{nor}{NOR}{normal subjects}
\newacronym{minf}{MINF}{previous myocardial infarction}
\newacronym{dcm}{DCM}{dilated cardiomyopathy}
\newacronym{hcm}{HCM}{hypertrophic cardiomyopathy}
\newacronym{arv}{aRV}{abnormal right ventricle}

\newacronym{rv}{RV}{Right Ventricle}
\newacronym{lv}{LV}{Left Ventricle}
\newacronym{vae}{VAE}{Variational Autoencoder}
\newacronym{sivae}{SIVAE}{Soft Introspective Variational Autoencoder}
\newacronym{arsivae}{AR-SIVAE}{Attribute Regularized Soft Introspective Variational Autoencoder}
\newacronym{gan}{GANs}{Generative Adversarial Networks}
\newacronym{mse}{MSE}{Mean Squared Error}
\newacronym{lvedv}{LVEDV}{Left Ventricular End-Diastole Volume}
\newacronym{myoedv}{MEDV}{Myocardial End-Diastole Volume}
\newacronym{rvedv}{RVEDV}{Right Ventricular End-diastole Volume}
\newacronym{ssim}{SSIM}{Structural Similarity Index Measure}
\newacronym{psnr}{PSNR}{Peak Signal-to-Noise Ratio}
\newacronym{sap}{SAP}{Separated Attribute Predictability}
\newacronym{scc}{SCC}{Spearman Correlation Coefficient}
\newacronym{mlp}{MLP}{Multi Layer Perceptron}
\newacronym{xai}{XAI}{explainable AI}
\newacronym{elbo}{ELBO}{Evidence Lower Bound}
\newacronym{lpips}{LPIPS}{Learned Perceptual Image Patch Similarity}

\title{Attribute Regularized Soft Introspective Variational Autoencoder for Interpretable Cardiac Disease Classification}
\name{Author(s) Name(s)\thanks{Some author footnote.}}
\address{Author Affiliation(s)}

\name{Maxime Di Folco $^{1}$ \qquad Cosmin I. Bercea$^{1,2,3}$\qquad Julia A. Schnabel$^{1,2,3,4}$}
\address{$^1$ Institute of Machine Learning in Biomedical Imaging, Helmholtz  Munich  \\
    {$^2$} Helmholtz AI, Helmholtz  Munich\\
    {$^3$} School of Computation, Information and Technology, Technical University of Munich\\
    {$^4$} School of Biomedical Engineering and Imaging Sciences, King's College London}

\begin{document}
%
\maketitle
\begin{abstract}

Interpretability is essential in medical imaging to ensure that clinicians can comprehend and trust artificial intelligence models. In this paper, we propose a novel interpretable approach that combines attribute regularization of the latent space within the framework of an adversarially trained variational autoencoder. Comparative experiments on a cardiac MRI dataset demonstrate the ability of the proposed method to address blurry reconstruction issues of variational autoencoder methods and improve latent space interpretability. Additionally, our analysis of a downstream task reveals that the classification of cardiac disease using the regularized latent space heavily relies on attribute regularized dimensions, demonstrating great interpretability by connecting the used attributes for prediction with clinical observations.
\end{abstract}
\section{Introduction}


Interpretability is crucial for transparent AI systems in medical imaging to build clinician trust and advance AI adoption in clinical workflows. As highlighted by Rudin \cite{Rudin:2019}, it is essential that models are inherently interpretable and not applied to black-box models to ensure their relevance.
Latent representation models like \gls{vae} have emerged as potent tools capable of encoding crucial hidden variables within input data \cite{higgins:2017, biffi:2020, Liu:2020}. Especially when dealing with data that contains different interpretable features (data attributes), supervised techniques can encode those attributes in the latent space. \cite{Mirza:2014, Fader:2017,engel2017latent,hadjeres:2017, Pati:2021}. In this context, Pati et al. \cite{Pati:2021} introduced an attribute-regularized method based on VAEs that aims to regularize each attribute, added as extra input, along a dimension of the latent space and, therefore, increase the latent space interpretability. Notably, Cetin et al. \cite{Cetin:2023} applied this architecture for cardiac attributes on MRI data, demonstrating a significant improvement in the interpretability of the latent representation and its relevance for a downstream classification task of cardiac disease. Nevertheless, VAE based methods may suffer from blurry reconstruction and could be problematic for any downstream task. In order to overcome this limitation while preserving the latent interpretability of VAEs, Daniel et al.~\cite{Daniel:2021} introduced a novel approach called \gls{sivae}. SIVAEs leverage the benefits of  VAEs and \gls{gan} by incorporating an adversarial loss into VAE training. In contrast to earlier methods that used additional discriminator networks ~\cite{pidhorskyi2018generative}, SIVAE utilizes the encoder and decoder of VAE in an adversarial manner.

The contributions of this paper are twofold. 
\begin{itemize}
     \item We propose the Attributed-regularized Soft Introspective Variational Autoencoder (AR-SIVAE) by combining an attribute regularization loss in the \gls{sivae} framework to keep the interpretability of the latent space while having better image generation capabilities. To the best of our knowledge, we are the first to introduce this loss in an adversarially trained \gls{vae}. We compare our method to the one described in Cetin et al. \cite{Cetin:2023} on a cardiac MRI public dataset. Our method mitigates the limitations associated with blurry reconstruction and slightly improves latent space interpretability performance.
     \item We further analyze the interpretability of the latent space for a downstream multi-class classification task. We demonstrate that the classification primarily relies on regularized dimensions and established correlations between the attributes used for classification and clinical observations.
\end{itemize} 

\section{Methods}

\begin{figure}[t]
    \centering
    \includegraphics[width=0.46\textwidth]{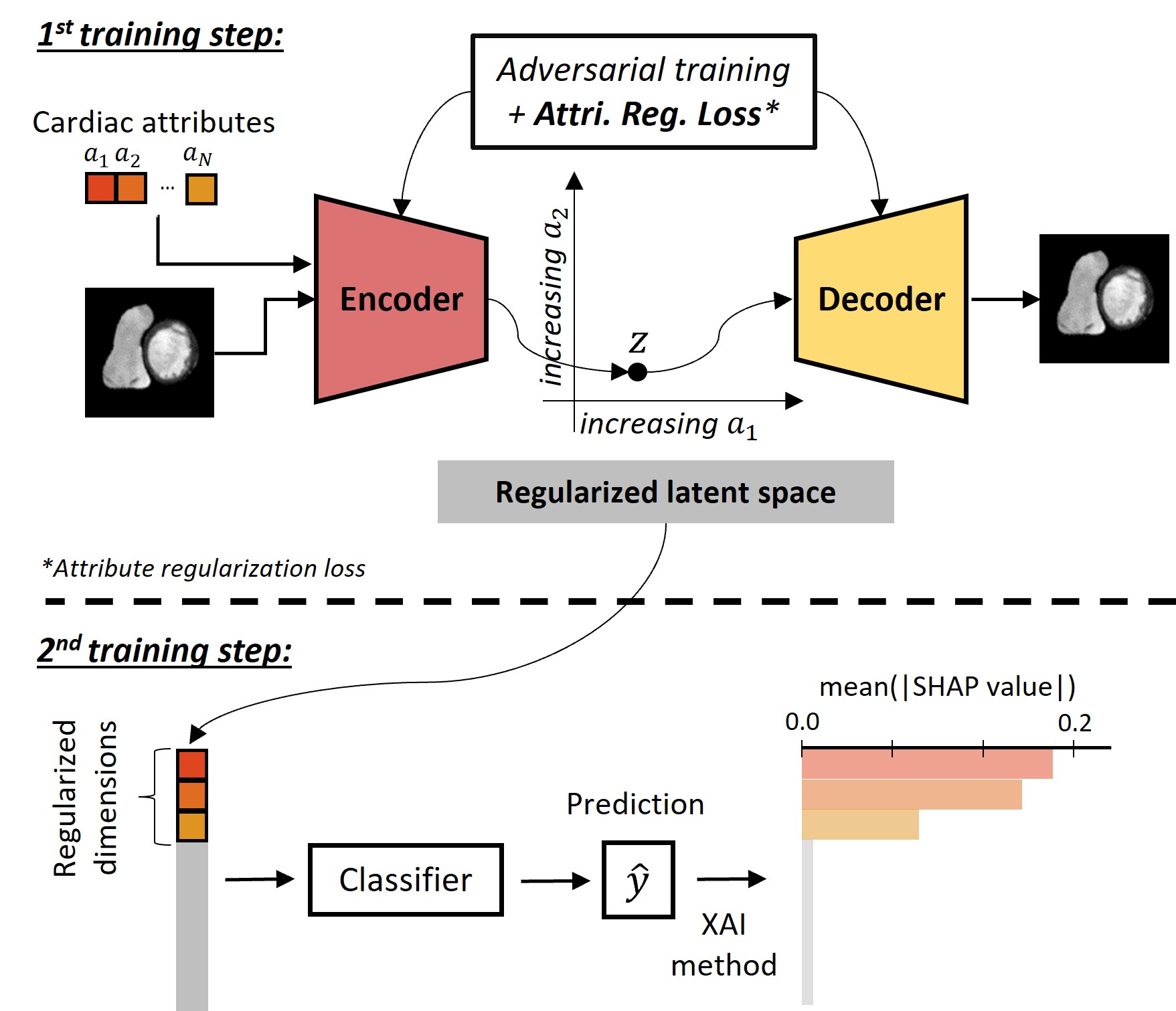}
    \caption{AR-SIVAE overview: Our training process first learns latent space regularised by cardiac attributes to then use it for a supervised downstream classification tasks (binary or multi-class). At inference, a new sample is projected onto the regularized latent space and then classified.}
    \label{fig:framework}
\end{figure}


\subsection{Preliminaries:}
\label{sec:rec}

\subsubsection{Attributed regularized VAE (Attri-VAE)}
Pati et al. \cite{Pati:2021} introduced an attributed-based regularization \gls{vae}, based on $\beta$-\gls{vae} \cite{higgins:2017}. This method aims to encode an attribute $a$ along a dimension $k$ of a $\mathbb{D}$-dimensional latent space
 \begin{math}\mathbf{z}: {z^k}, k \in [0,\mathbb{D}) \end{math}, such that the attribute value $a$ of the generated data increases when we traverse along $k$. The attribute regularization is based on an attribute distance matrix $D_a$ and a similar distance matrix $D_k$ computed from the regularized dimension $k$. They are defined as follows:

\begin{equation}
	\centering
	D_a(i,j) = a(x_i) - a(x_j) \text{;  } D_k(i,j) = z_i^k - z_j^k
\end{equation}

where $\mathbf{x_i}, \mathbf{x_j} \in \mathbb{R^N}$ are two high-dimensional samples of dimensions $N$ (with $N >> \mathbb{D}$). 
The attribute regularization loss term is then computed as follows and added to the $\beta$-VAE loss:

\begin{equation}
	\centering
	L_{k,a} = \gamma_{reg} \times MAE(tanh(\delta D_k) - sgn(D_a))
	\label{eq:attr}
\end{equation}

where $MAE(.)$ is the mean absolute error, $tanh(.)$ is the hyperbole tangent function, $sgn(.)$ is the sign function, $\delta$ and $\gamma_{reg}$ are tunable hyperparameters which decide the spread of the posterior distribution and weight the importance given to the regularization, respectively.
\subsubsection{Soft Introspective Variational Autoencoder (SIVAE)}

The \gls*{sivae} framework proposed by \cite{Daniel:2021} is an adversarially trained \gls{vae}. Its encoder is trained to distinguish between real and generated samples by minimizing the KL divergence between the latent distribution of real samples and the prior while maximizing the KL divergence of generated samples. Conversely, the decoder aims to deceive the encoder by reconstructing real data samples using the standard \gls{elbo} and minimizing the KL divergence of generated samples embedded by the encoder. The optimization objectives for the encoder, $E_{\Phi}$, and decoder, $D_{\theta}$, to be maximized are formulated as follows:
\begin{flalign}
	\label{eq::sivae}
	& \mathcal{L}_{E_{\phi}}(\mathbf{x},\mathbf{z}) = ELBO(\mathbf{x}) - \frac{1}{\alpha}(exp(\alpha ELBO(D_\theta(\mathbf{z})),\\
	& \mathcal{L}_{D_{\theta}}(\mathbf{x},\mathbf{z}) = ELBO(\mathbf{\mathbf{x}}) + \gamma ELBO(D_\theta(\mathbf{z})) \nonumber
\end{flalign}
where $\alpha \geq 0$ and $\gamma \geq 0$ are hyper-parameters.

\subsection{Proposed method: AR-SIVAE}
\label{sec:ours}

We propose in this work an \gls{arsivae} by adding the attribute regularization loss defined in Eq. \ref{eq:attr} to the encoder loss of the \gls{vae} (Fig. \ref{fig:framework}). The optimization objective of the encoder becomes: 
\begin{flalign}
	\label{eq:OURS}
	\mathcal{L}_{E_{\phi}}(\mathbf{x},\mathbf{z}) = & ELBO(\mathbf{z}) - \frac{1}{\alpha}(exp(\alpha ELBO(D_\theta(\mathbf{z})) \\
     & + \gamma_{reg} \times L_{r,a} \nonumber
\end{flalign}

where $\gamma_{reg}$ a hyperparameter that weights the attribute regularization loss term. 

The training process of our framework is divided into two parts, illustrated in Fig. \ref{fig:framework}. The first step learns a regularized latent space by adding an attribute regularization loss to the \gls{sivae} framework. Then, in a second step, a classifier based on \gls{mlp} architecture, is trained in a supervised way on the regularized latent space to classify cardiac disease. At inference, the encoder and classifier are frozen and used to project a new sample onto the regularized latent space and predict the disease, respectively. As a post-hoc explanation method, we use Shapley values \cite{Sundararajan:2017} to attribute the contributions of individual features, i.e. latent dimensions, to the model's predictions, highlighting the factors influencing cardiac disease classification. 



\section{Experiments and results}

We compare the proposed method \gls{arsivae} through a set of experiments to evaluate the reconstruction performance and the interpretability of the learned representation against: $\beta$-VAE, SIVAE and Attri-VAE. $\beta$-VAE and SIVAE methods do not include attribute regularization and will be used as baselines for the reconstruction performance. Attri-VAE \cite{Pati:2021} is considered as reference for attribute regularization and interpretability of the latent space. Implementation details of each method and the code are available here: \url{https://github.com/maxdifolco/AR-SIVAE}.

\textit{Data and preprocessing steps:} We use the cardiac MRI public dataset: ACDC \cite{Bernard:2018}, which contains 150 \gls{mr} images of 5 different classes: \gls{nor} and patients with either \gls{minf}, \gls{dcm}, \gls{hcm} or \gls{arv}. From the ground-truth segmentation at end-diastole,  we select the basal slice of each volume and then compute the cardiac volumes of each of the regions of interest: \gls{lvedv}, \gls{myoedv}, and \gls{rvedv}. An extensive study of the attribute choice is described in \cite{Cetin:2023}.  We choose a limited number of attributes in this paper to facilitate the interpretability analysis of the predictions. 

\subsection{Reconstruction}

\begin{figure}[ht]
    \centering
    \includegraphics[width=0.36\textwidth]{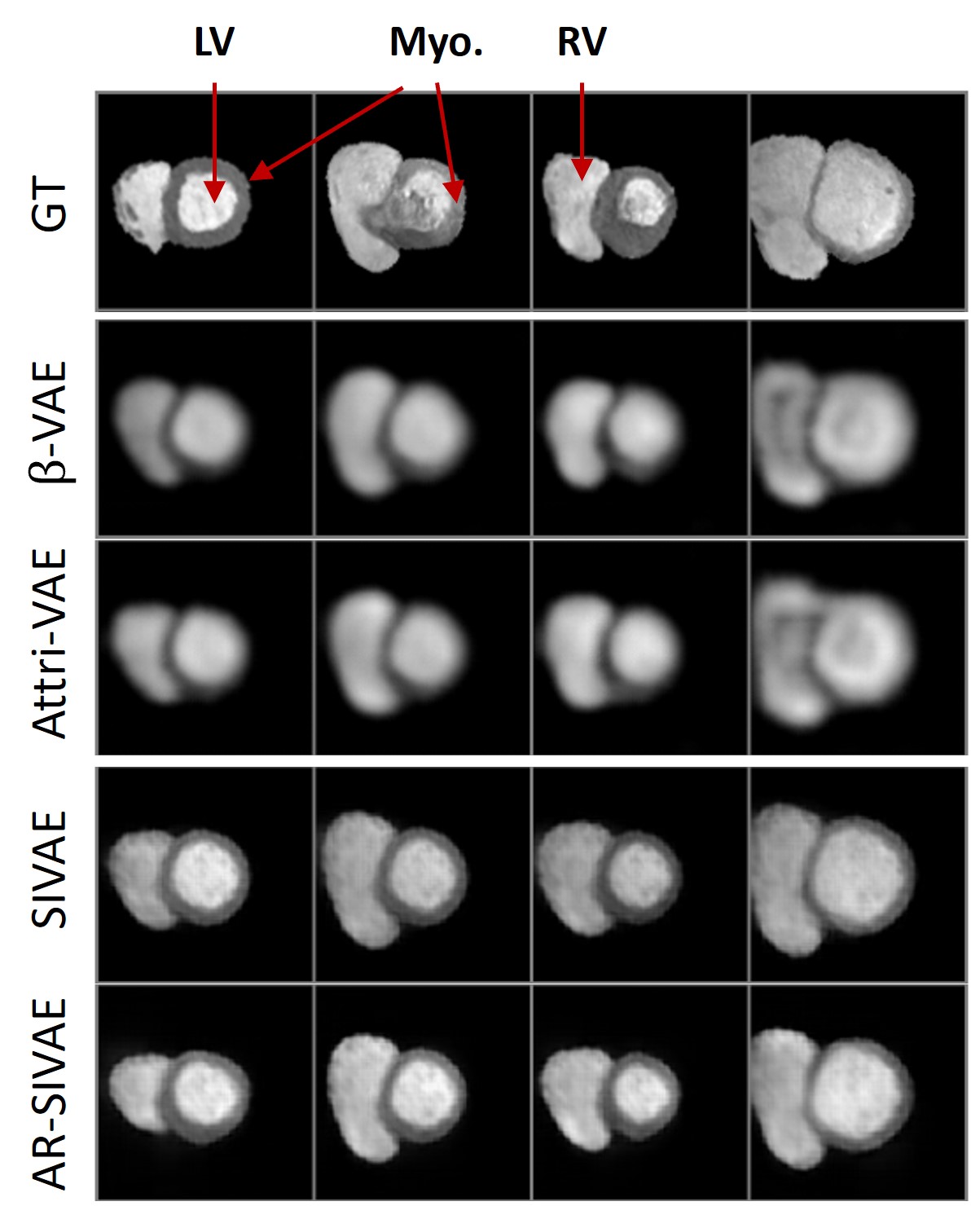}
    \caption{Qualitative evaluation of the reconstruction performance for $\beta$-VAE, \gls{sivae}, Attri-VAE and AR-SIVAE. The first line corresponds to the Ground Truth (GT). The red arrows show the main cardiac anatomies element: Left Ventricle (LV), Myocardium (Myo.) and Right Ventricle (RV). }
    \label{fig:rec_samples}
\end{figure}

We initially assess the reconstruction performance of the compared methods both qualitatively, as illustrated in Fig. \ref{fig:rec_samples}, and quantitatively, as detailed in Table. \ref{tab:rec_performance}. The evaluation employs metrics such as \gls{psnr}, \gls{ssim}, and \gls{lpips} \cite{zhang2018unreasonable}. Despite similar performance for all the compared methods in terms of \gls{psnr} and \gls{ssim} metrics in Table. \ref{tab:rec_performance}, Fig. \ref{fig:rec_samples} shows the ability of the SIVAE based methods to reconstruct non-blurry samples contrary to $\beta$-VAE and Attri-VAE. However, SIVAE based reconstructions display more subtle differences from the ground truth in the global shape, particularly for the \gls{rv}, which contribute to lower the \gls{psnr} and \gls{ssim} scores. While they are widely used metrics to assess the similarity between two images, \gls{psnr} and \gls{ssim} often fail to detect nuances of human perception \cite{zhang2018unreasonable}. We also employ the \gls{lpips} metric to address this limitation, revealing a significant improvement for SIVAE-based methods. Furthermore, both kinds of methods ($\beta$-VAE based or SIVAE based) achieve similar performance with and without attribute regularization, suggesting that the addition of the regularization term has minimal influence on reconstruction quality (except for the \gls{ssim} metric for SIVAE based methods). We also observe in Fig. \ref{fig:rec_samples} that all of the methods fail to reconstruct the myocardial thickness, and due to the blurriness of the reconstruction, $\beta$-VAE and Attri-VAE delimit the myocardium to the left ventricle cavity poorly.

\begin{table}[ht]
	\centering
	\resizebox{0.44\textwidth}{!}{
\begin{tabular}{l|c|ccc}
		\toprule
		
		& \textbf{Reg.} &  \textbf{SSIM} $\uparrow$&  \textbf{ PSNR } $\uparrow$ & \textbf{LPIPS} $\downarrow$ \\
		\midrule
		
		$\beta$-VAE & \xmark & 0.74 $\pm$ 0.04  & \textbf{20.8 $\pm$ 1.4} & 0.24 $\pm$ 0.03	 \\
		Attri-VAE  & \cmark & 0.73 $\pm$	0.04 & 20.7 $\pm$ 1.5 & 0.25 $\pm$ 0.03 \\
	\midrule
		SIVAE & \xmark & \textbf{0.76 $\pm$ 0.04} & 20.4 $\pm$ 1.6 & \textbf{0.09 $\pm$ 0.02} \\
		\textbf{AR-SIVAE} (ours) & \cmark & 0.71 $\pm$ 0.05 & 20.6 $\pm$ 1.7 & \textbf{0.09 $\pm$ 0.02} \\
		\bottomrule
\end{tabular}}

	\caption{Evaluation in terms of reconstruction performance using \gls{ssim}, \gls{psnr} and \gls{lpips} metrics of the compared methods with or without attributed regularization (Reg. column).}
	\label{tab:rec_performance}
\end{table}

\subsection{Interpretability of the latent space}

\begin{table}[ht]
	\centering
	\resizebox{0.42\textwidth}{!}{
\begin{tabular}{l|c|cccc}
		\toprule
		
		& \textbf{ Reg.} &  \textbf{ Interp.}  & \multicolumn{1}{c}{  \textbf{ SCC. }  } & \multicolumn{1}{c}{  \textbf{ Mod. }  } &  \multicolumn{1}{c}{  \textbf{ SAP }  } \\ 
            \midrule
	
		$\beta$-VAE & \xmark & 0.21  & 0.65	&  0.79  &  0.01   \\
            Attri-VAE & \cmark &  0.78  & 0.89 	& \textbf{0.82}  & 0.51 \\		
            
		\midrule

		  SIVAE & \xmark & 0.42 	& 0.71 	&  0.81 &  0.04 \\
		\textbf{AR-SIVAE} \textbf{} & \cmark & \textbf{0.83} 	& \textbf{0.91} 	&        \textbf{0.82}  &  \textbf{0.59}  \\
	\bottomrule
        
\end{tabular}
}

	\caption{Comparison of disentanglement performance metrics. All the metrics are between 0 and 1, with 1 being the best performance.}
	\label{tab:interp_performance}
\end{table}

\begin{figure}[ht]
    \centering
    \subfloat[Attri-VAE]{\includegraphics[width=0.42\linewidth]{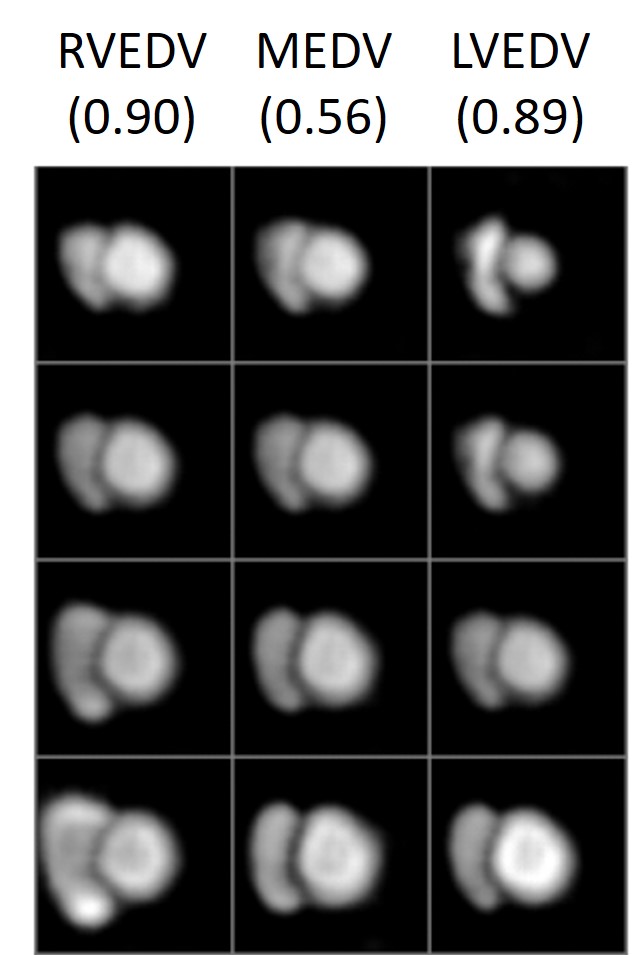}
    	\label{fig:ld_VAE}}
    \subfloat[AR-SIVAE]{\includegraphics[width=0.42\linewidth]{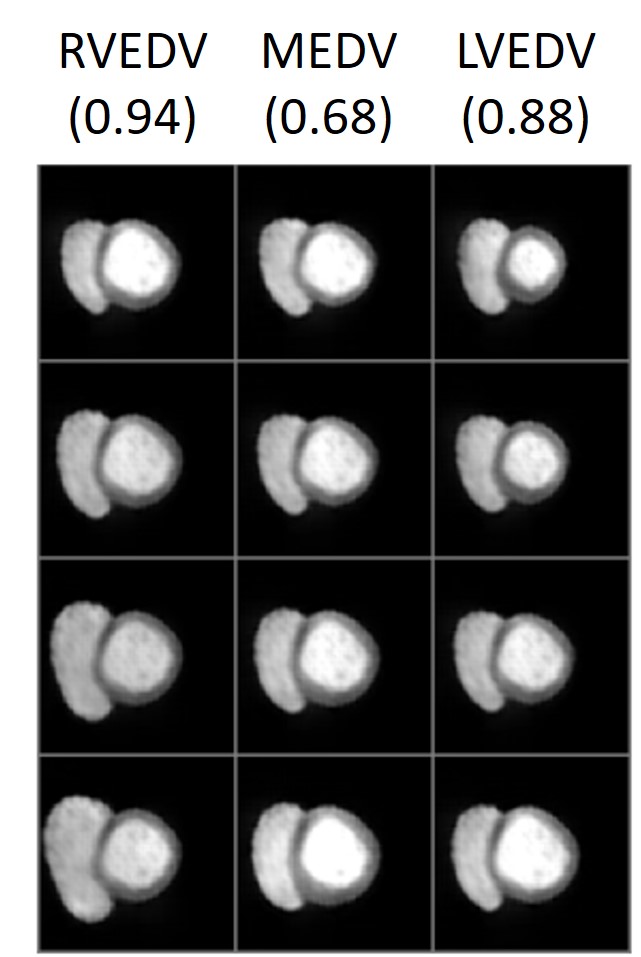}
    	\label{fig:ld_SIVAE}}
    \caption{Walk in the regularized latent dimensions for (a) Attri-VAE and (b) AR-SIVAE. The corresponding cardiac attribute with the Interpretability score in parenthesis is displayed above each dimension.}
    \label{fig:latent_dim}
\end{figure}

To evaluate the disentanglement of the latent space, we report in Table. \ref{tab:interp_performance} the \textit{Interpretability} metric \cite{Abdel:2018}, the \textit{Modularity} metric \cite{Ridgeway:2018}, the \gls{sap} and the \gls{scc}. An improvement across all interpretability metrics is evident when comparing the method's results with and without attribute regularization. The proposed method demonstrates superior performance against Attri-VAE for the mean Interpretability score and the \gls{sap} metrics. The difference in the Interpretability score is primarily due to a lower Interpretability score for the myocardium dimension (score for each regularized dimension is reported in Fig. \ref{fig:latent_dim}). This discrepancy can be attributed to Attri-VAE's limited capability in reconstructing the myocardium, as demonstrated in the preceding section. Fig. \ref{fig:latent_dim} also illustrates a walk along each regularized dimension. The evolution of the attributes is easily noticeable for the \gls{lvedv} and \gls{rvedv}, unlike the \gls{myoedv} where the evolution is more subtle.

\subsection{Interpretable classification}

\begin{figure}[h!]
    \centering
    \includegraphics[width=0.48\textwidth]{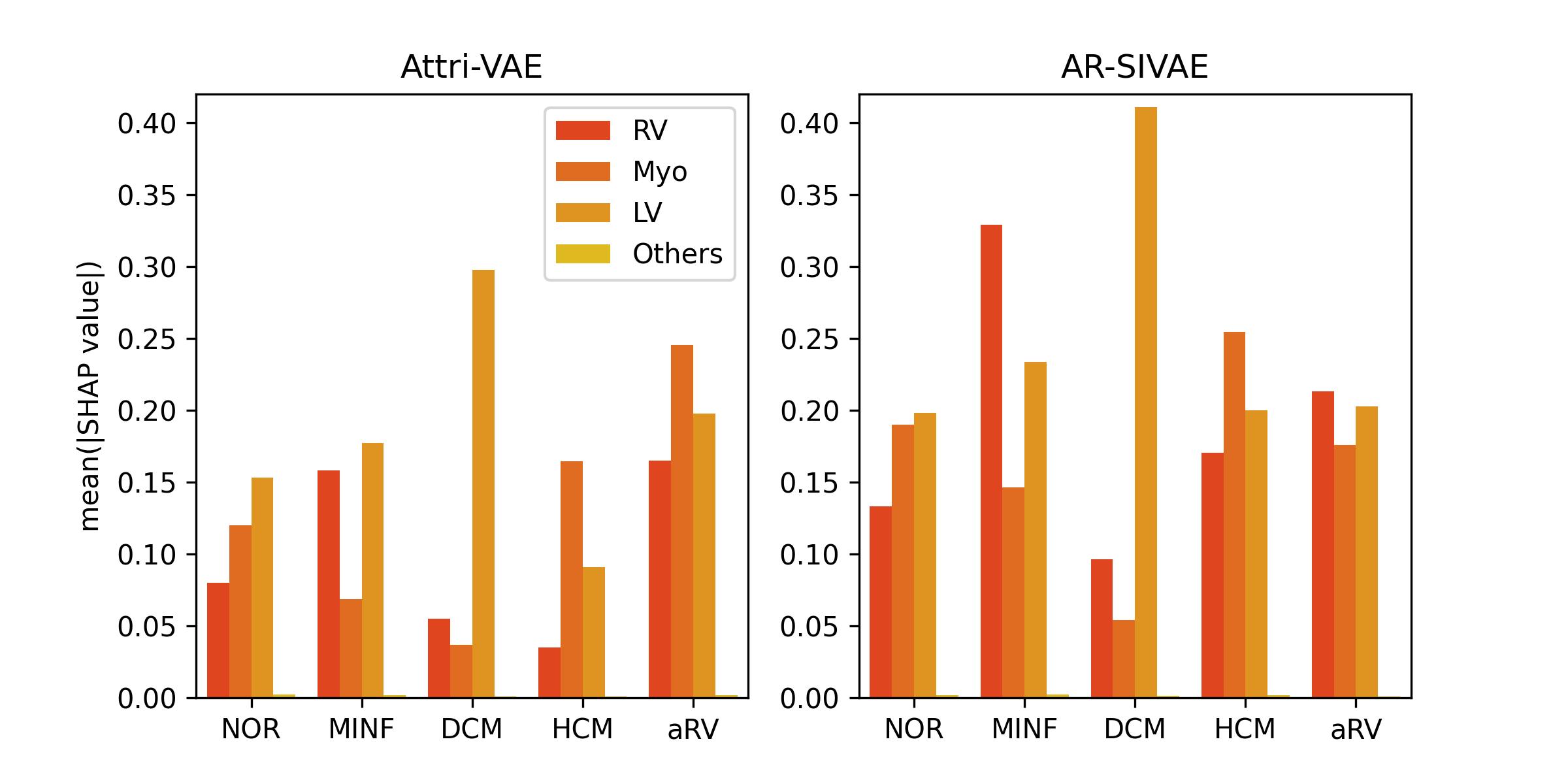}
    \caption{Average impact on model outputs magnitude of each dimension using SHAP values for multi-class classification of Normal subjects (NOR) and patients with previous myocardial infarction (MINF), dilated cardiomyopathy (DCM), hypertrophic cardiomyopathy (HCM) and abnormal right ventricle (aRV). "Others" correspond to all non-regularised dimensions' aggregated mean($|$SHAP value$|$).}
    \label{fig:interp}
\end{figure}

The results for the downstream classification are reported in Table 3. The baseline is obtained by training a Support Vector Machine (SVM) with a Radial Basis Function (RBF) kernel and represents the predictive capabilities of the three attributes. For binary and multi-class tasks, the attribute regularized methods show an improvement in the performance and reach almost the baseline performance. Fig. \ref{fig:interp} shows the average impact on model outputs magnitude of each dimension and each class with "Others" corresponding to the aggregated contributions for non-regularized dimensions. We initially observed that both methods use only the three regularized dimensions for predictions. For both methods, the predictions of the \gls{hcm} and \gls{dcm} classes mainly rely on the myocardium dimension and the \gls{lv} dimension, respectively, aligning with the characteristics described in the ACDC dataset for each disease: \gls{hcm} patients exhibit higher cardiac mass and a thicker myocardium, resulting in a larger myocardium volume, while \gls{dcm} patients have a high \gls{lvedv}. Regarding the \gls{arv} class, characterized by an abnormal \gls{rvedv}, Attri-VAE primarily utilized the \gls{lv} and myocardium dimensions, in contrast to our method where the \gls{rv} and myocardium dimensions contribute the most. These observations underscore the interpretability of our method. However, basing the predictions on cardiac attributes for interpretability comes with the cost of lower classification performance (e.g. using non-interpretable methods, the original dataset paper \cite{Bernard:2018} reports an accuracy up to 0.91 for the same multi-class classification task).

\begin{table}[t]
	\centering
        \resizebox{0.38\textwidth}{!}{
\begin{tabular}{l|c|ccc}
		\toprule
		
		& \textbf{Reg.} & \textbf{Acc.}  & \multicolumn{1}{c}{  \textbf{F1-score}} & \multicolumn{1}{c}{  \textbf{AUROC}}\\
		\midrule
            \rowcolor{Gray}
            \multicolumn{5}{c}{\textbf{Binary classification}} \\
            \midrule
           $\beta$-VAE & \xmark & 0.68 & 0.60 & 0.65 \\
            Attri-VAE & \cmark & \textbf{0.80} & \textbf{0.73} & \textbf{0.76}  \\
            
            SIVAE & \xmark & 0.68 & 0.58 & 0.61 \\
            \textbf{AR-SIVAE} (ours) & \cmark & \textbf{0.80} & 0.71 & 0.73  \\
            \hline
            SVM (baseline) & - & 0.80 & 0.75 & 0.84   \\
		\midrule
            \rowcolor{Gray}
            \multicolumn{5}{c}{\textbf{Multi-class classification}} \\
            \midrule
            $\beta$-VAE & \xmark & 0.36 & 0.32 & 0.71 \\
            Attri-VAE & \cmark & 0.54 & 0.55  & \textbf{0.86} \\
            
            SIVAE & \xmark & 0.46 & 0.43 & 0.71 \\
           \textbf{AR-SIVAE} (ours) & \cmark & \textbf{0.58} & \textbf{0.58} & 0.85  \\
            \hline
           SVM (baseline) & - &  0.6 & 0.63  & 0.88  \\
	    \bottomrule
\end{tabular}
}

	\caption{Evaluation of the downstream classification tasks for binary and multi-class classification. The compared methods aim to achieve, using only the input image, the performance of the baseline that represents the predictive capabilities of the three attributes alone. The baseline is obtained using a Support Vector Machine (SVM) with a Radial Basis Function (RBF) kernel on only the attributes.}
	\label{tab:class_performance}
\end{table}

\section{Conclusion}

This paper introduces the Attribute regularized Soft Introspective Variational Autoencoder (AR-SIVAE), which combines attribute regularization within the SIVAE framework to enhance the interpretability of the latent space while improving image generation capabilities. Moreover, our analysis reveals that the classification process relies predominantly on regularized dimensions and achieves great interpretability by connecting the attributes used for classification and clinical observations. Future work will focus on improving our method's image reconstruction and classification performance, as it adds interpretability but does not yet achieve state-of-the-art results in the latter aspect.


\section{Compliance with ethical standards}

This research study was conducted retrospectively using human subject data made available in open access by Bernard et al. \cite{Bernard:2018}. Ethical approval was not required, as confirmed by the license attached with the open access data.

\bibliographystyle{IEEEbib}
\bibliography{strings,refs}

\end{document}